\documentclass[a4paper,11pt]{article}

\pdfoutput=1 
\usepackage{jinstpub} 

\usepackage{lmodern}															
\usepackage{isotope}															
\usepackage{upgreek}															
\usepackage{textcomp}															
\usepackage{booktabs}															
\usepackage{esint}
\usepackage[all]{hypcap}													
\usepackage{wasysym}															
\usepackage{lineno}																
\newcommand{\unitcm}{\,\text{cm}}									
\newcommand{\unitmm}{\,\text{mm}}									
\newcommand{\unitmum}{\,\text{{\textmu}m}}				
\newcommand{\unitnm}{\,\text{nm}}									
\newcommand{\unitmsq}{\,\text{m}^2}								
\newcommand{\unitkeV}{\,\text{keV}}								
\newcommand{\unityr}{\,\text{yr}}									
\newcommand{\unitkHz}{\,\text{kHz}}								
\newcommand{\unitdeg}{^{\circ}}										
\newcommand{\unitpdeg}{\,^1\!\text{/}_\text{deg}}	
\newcommand{\unitK}{\,\text{K}}										
\newcommand{\unitC}{\,^{\circ}\text{C}}						
\newcommand{\unitbar}{\,\text{bar}}								
\newcommand{\unitmbar}{\,\text{mbar}}							

\title{\boldmath Reflectivity of VUV-sensitive Silicon Photomultipliers in Liquid Xenon}

\author[1,a]{M.~Wagenpfeil \note[a]{Corresponding author}}
\author[1]{T.~Ziegler}
\author[1]{J.~Schneider}
\author[2, b]{A.~Fieguth \note[b]{Now at Physics Department, Stanford University}}
\author[2]{M.~Murra}
\author[2]{D.~Schulte}
\author[2]{L.~Althueser}
\author[2]{C.~Huhmann}
\author[2]{C.~Weinheimer}
\author[1]{T.~Michel}
% ...

\affiliation[1]{Erlangen Centre for Astroparticle Physics (ECAP), Friedrich-Alexander University Erlangen-N\"urnberg, Erlangen 91058, Germany}
\affiliation[2]{Institut f\"ur Kernphysik, Westf\"alische Wilhelms-Universit\"at M\"unster, 48149 M\"unster, Germany}
% ...

\emailAdd{michael.wagenpfeil@fau.de}

\abstract{%
Silicon photomultipliers are regarded as a very promising technology for next-generation, cutting-edge detectors for low-background experiments in particle physics. This work presents systematic reflectivity studies of Silicon Photomultipliers~(SiPM) and other samples in liquid xenon at vacuum ultraviolet~(VUV) wavelengths. A dedicated setup at the University of Münster has been used that allows to acquire angle-resolved reflection measurements of various samples immersed in liquid xenon with $0.45\unitdeg$ angular resolution. Four samples are investigated in this work: one Hamamatsu VUV4 SiPM, one FBK VUV-HD SiPM, one FBK wafer sample and one Large-Area Avalanche Photodiode~(LA-APD) from EXO-200. The reflectivity is determined to be $25\text{--}36\,\%$ at an angle of incidence of $20\unitdeg$ for the four samples and increases to up to $65\,\%$ at $70\unitdeg$ for the LA-APD and the FBK samples. The Hamamatsu VUV4 SiPM shows a decline with increasing angle of incidence. The reflectivity results will be incorporated in upcoming light response simulations of the nEXO detector.
}


\arxivnumber{2104.07997}

\begin{document}

\maketitle
\flushbottom

\section{Introduction}\label{sec:Intro}

Neutrinoless double beta ($0\upnu\upbeta\upbeta$) decay is a hypothetical, second-order weak transition of a nucleus that involves the decay of two neutrons into two protons and two electrons without the emission of antineutrinos. Observing $0\upnu\upbeta\upbeta$-decay would require new physics beyond the Standard Model of particle physics, e.g.~lepton number violation~\cite{schechter82,gomez12}.

Experimental searches for $0\upnu\upbeta\upbeta$-decay are complicated by a half-life longer than ${\mathcal{O}}\left(10^{25}\,\text{yr}\right)$ \cite{EXO19-limit,CUORE20,GERDA20,KamLAND16}. nEXO is a planned, tonne-scale $0\upnu\upbeta\upbeta$-decay search experiment with a cylindrical, single-phase time projection chamber~\cite{nEXO18-pCDR} and the next-generation successor of the successful EXO-200 experiment~\cite{EXO11-design, EXO11-obs}. The target energy resolution is $\sigma$/$Q_{\upbeta\upbeta}=1\,\%$ at the $Q$-value of the \isotope[136]{Xe}-$0\upnu\upbeta\upbeta$-decay ($2458.10{\pm}0.31\unitkeV$~\cite{redshaw07,mccowan10}) and the half-life sensitivity is $T_{1\text{/}2}^{0\upnu} > 9{\times}10^{27}\unityr$ at $90\,\%$ C.L. after $10$ years of data taking~\cite{nEXO17-sens}.

The detector will be equipped with a $4.5\unitmsq$ array of silicon photomultipliers~(SiPMs) for the detection of the liquid xenon~(LXe) scintillation light~\cite{nEXO17-tiles}. The SiPMs cover the entire inner wall of the time projection chamber between the inner cryostat and the field cage~\cite{nEXO18-pCDR,nEXO18-field}. SiPMs are semiconductor photosensors with single-photon resolution~\cite{otte06}, offer a compact and robust geometry, high gain, low bias voltage, easy scalability, and can be manufactured with high radiopurity materials~\cite{ostrovskiy15}, which makes them particularly suitable for low-background experiments.

The energy resolution for particle interactions in nEXO is a crucial parameter and depends on the capability of the detector to efficiently collect the scintillation light produced by particle interactions. The nEXO collaboration undertakes extended SiPM characterization efforts to find the optimal SiPM candidate for their LXe scintillation photon detector~\cite{ostrovskiy15,nEXO18-sipm,nEXO19-sipm}. Precise knowledge of the optical behaviour of the SiPMs in LXe is crucial to properly simulate events in the detector. These simulations have an impact on the event reconstruction efficiency and the predicted sensitivity of nEXO. In particular, knowledge of the reflectivity of the SiPMs, regarded as the photosensors of choice for nEXO, in LXe and at LXe scintillation light wavelengths (centred around $175\unitnm$~\cite{fujii15}) in the vacuum-ultraviolet~(VUV) is required to accurately model the light transport in the detector in optical simulations.

For some photosensor candidates, the VUV-reflectivity has been measured before by the nEXO collaboration in vacuum and in LXe. In~\cite{nEXO19-reflVac}, samples from the manufacturer FBK~(Fondazione Bruno Kessler) are examined in vacuum, with their reflectivity measured as a function of the angle of incidence and the wavelength. The optical parameters of the samples are determined via dedicated models and their behaviour in LXe is extrapolated. The reflectivity of a wafer sample is measured to vary between $35\,\%$ and $65\,\%$ for monochromatic light at $175\unitnm$. It was also calculated to \textasciitilde$\,53\,\%$ for the LXe scintillation spectrum and angles of incidence below $60\unitdeg$. A second study by the LIXO group~\cite{nEXO19-reflLXe}, which focused on Hamamatsu SiPMs in LXe, measuring both SiPM reflectivity and photon detection efficiency using a dedicated setup and LXe scintillation light. One SiPM is used as reflector while a second SiPM measures the reflected photon flux. The response of both SiPMs is read out directly. The reflectivity of a sample from the same wafer as above is in accordance with~\cite{nEXO19-reflVac} while for the Hamamatsu SiPM, it drops from $28\,\%$ to $20\,\%$ between $15\unitdeg$ and $65\unitdeg$.

In this work, the reflectivity is measured with a different approach using angle-resolved reflection measurements. This allows for the measurement of the dependence of the angular reflectivity on the angle of incidence in more detail than previous works. Additionally, the effect of the surface micro-structure on the reflection of SiPMs can be observed and investigated for the first time. The unusual shape of the Hamamatsu VUV4 SiPM reflectivity curve observed in~\cite{nEXO19-reflLXe} is verified. This work also reports results from an FBK VUV-HD SiPM, which has not yet been examined in LXe. In the following Sections, we first describe the experimental setup (Section~\ref{sec:hardware}) and sources of systematic uncertainties (Section~\ref{sec:systematics}). In Section~\ref{sec:results}, we present the measured results, which will be used in the upcoming optical simulations and sensitivity studies of nEXO~\cite{EXO16-light, nEXO17-sens}.

\section{Hardware and measurement approach}\label{sec:hardware}
\subsection{Scope of the reflectivity studies}\label{ssec:scope}

Reflection refers to the process by which electromagnetic flux, incident on a stationary surface or medium, leaves that surface or medium from the incident side without change in frequency~\cite{nicodemus77}. \textit{Reflectivity} describes the fraction of photons reflected by a layer with a thickness well above the wavelength of the reflected light. The reflectivity depends on the angle of incidence~(AOI) $\theta_{\text{i}}$ with respect to the normal of the reflecting surface. The angular reflectivity $R$ describes the fraction of photons reflected into a certain solid angle $\text{d}\Omega$ while the total reflectivity $R_{\text{tot}}$ corresponds to the total fraction of reflected photons. Reflection phenomena are typically a mixture of \textit{specular reflection}, where incoming light is reflected at the same angle as the incident angle, and \textit{diffuse reflection}, where the light is scattered at a large angle range following Lambert's cosine law~\cite{koppal14}.

SiPMs consist primarily of smooth surfaces without significant roughness. For this reason, diffuse reflectivity is typically negligible for SiPMs. The total reflectivity is then equal or very close to the specular reflectivity.

\subsection{Measurement setup}\label{ssec:setup}

\begin{figure}[t!]
	\centering
	\includegraphics[width=0.60\textwidth]{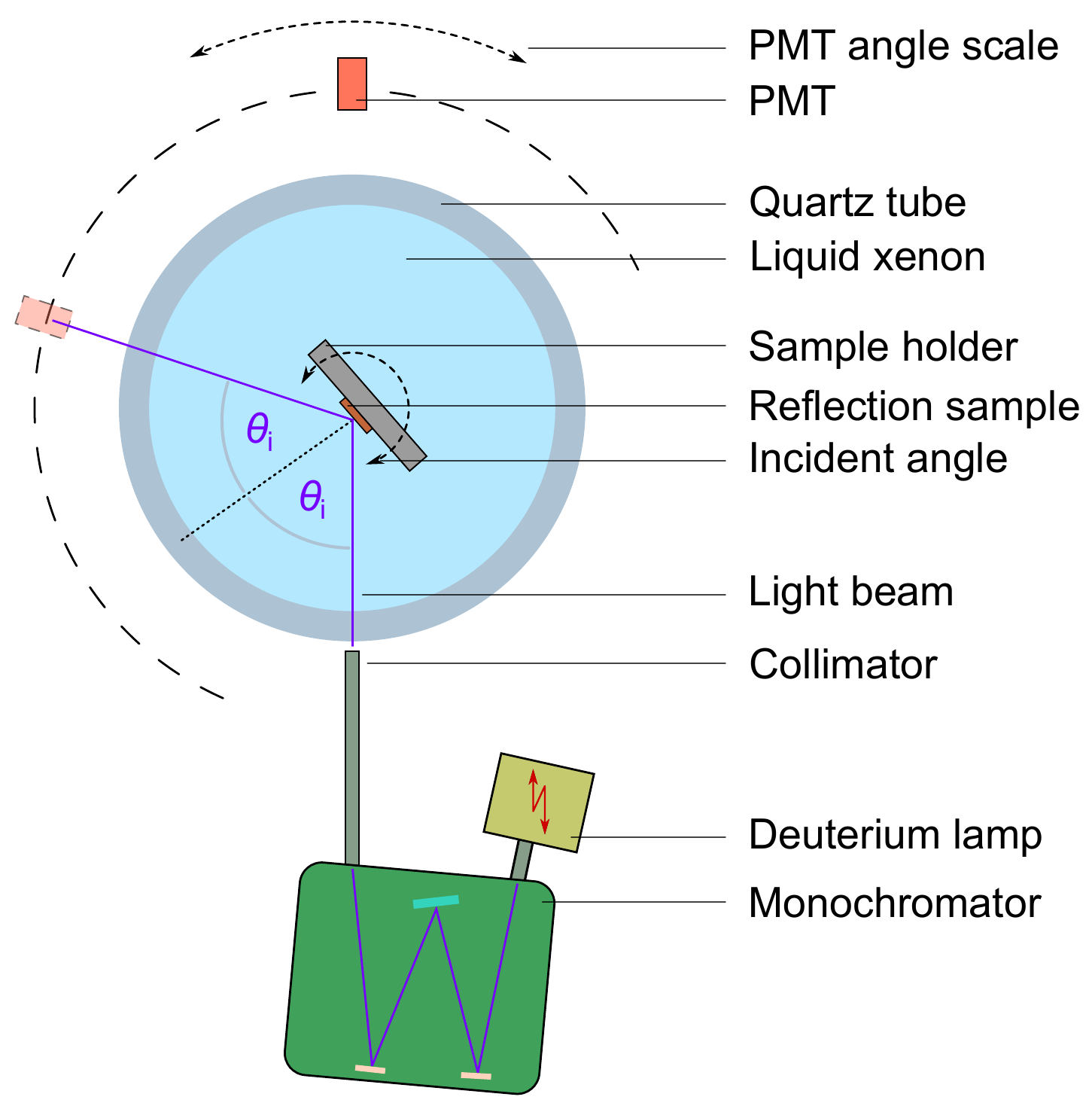}
	\caption[Scheme of the Münster reflection setup]{Schematic overview of the reflection setup at the University of Münster~\cite{bokeloh13,levy14,althueser20}. A light system produces a narrow VUV spectrum. The photons are guided into a quartz tube filled with LXe. The light is reflected by a sample located along the rotation axis of the tube. Reflected photons are detected by a photomultiplier tube~(PMT). The angle of incidence $\theta_{\text{i}}$ is set by rotating the quartz tube.}
	\label{fig:Refl-setup-scheme}
\end{figure}

\begin{figure}[t!]
	\centering
	\includegraphics[width=0.99\textwidth]{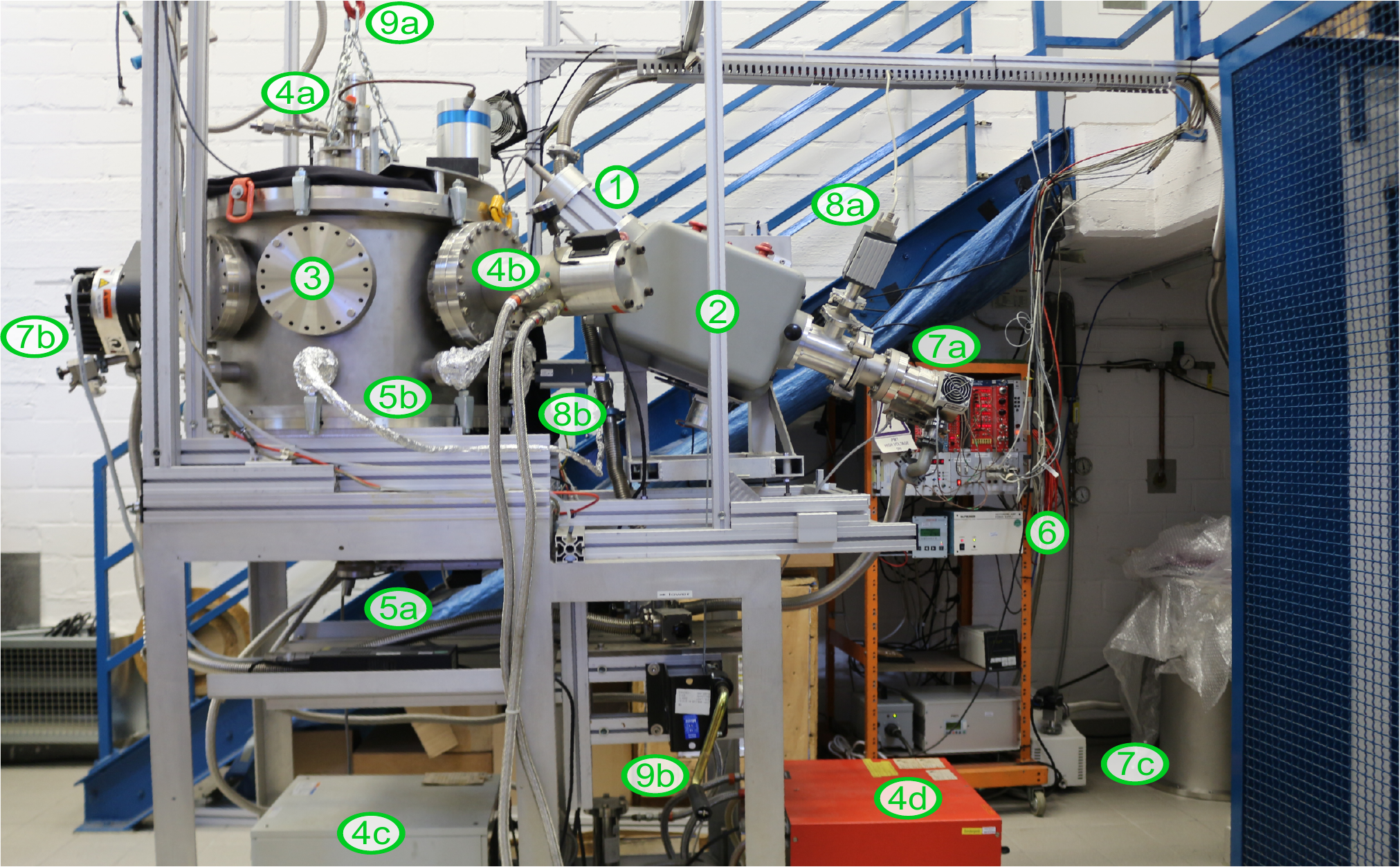}
	\caption[Picture of the Münster reflection setup]{Side view of the reflection setup at the University of Münster consisting of (1) the deuterium lamp, (2) the monochromator, the light guide (behind PMT cryohead), (3) the main chamber, (4a) the cryoheads for the xenon liquefaction and (4b) for PMT plus cold shield, (5a) the PMT stepping motor and (5b) the PMT signal and high voltage feedthrough as well as (6) the electronic readout, controls and DAQ interfaces. Compressors for the cryoheads are on the bottom (4c/d). Turbopumps for the monochromator (7a) and main chamber (7b) are connected to a common scroll pump (7c). Digital pressure sensors monitor both vacuum systems (8a/b). The quartz tube can be lifted via a pulley (9a) connected to a manual hoist (9b).}
	\label{fig:Refl-setup-picture}
\end{figure}

\begin{figure}[t!]
	\centering
	\includegraphics[width=0.99\textwidth]{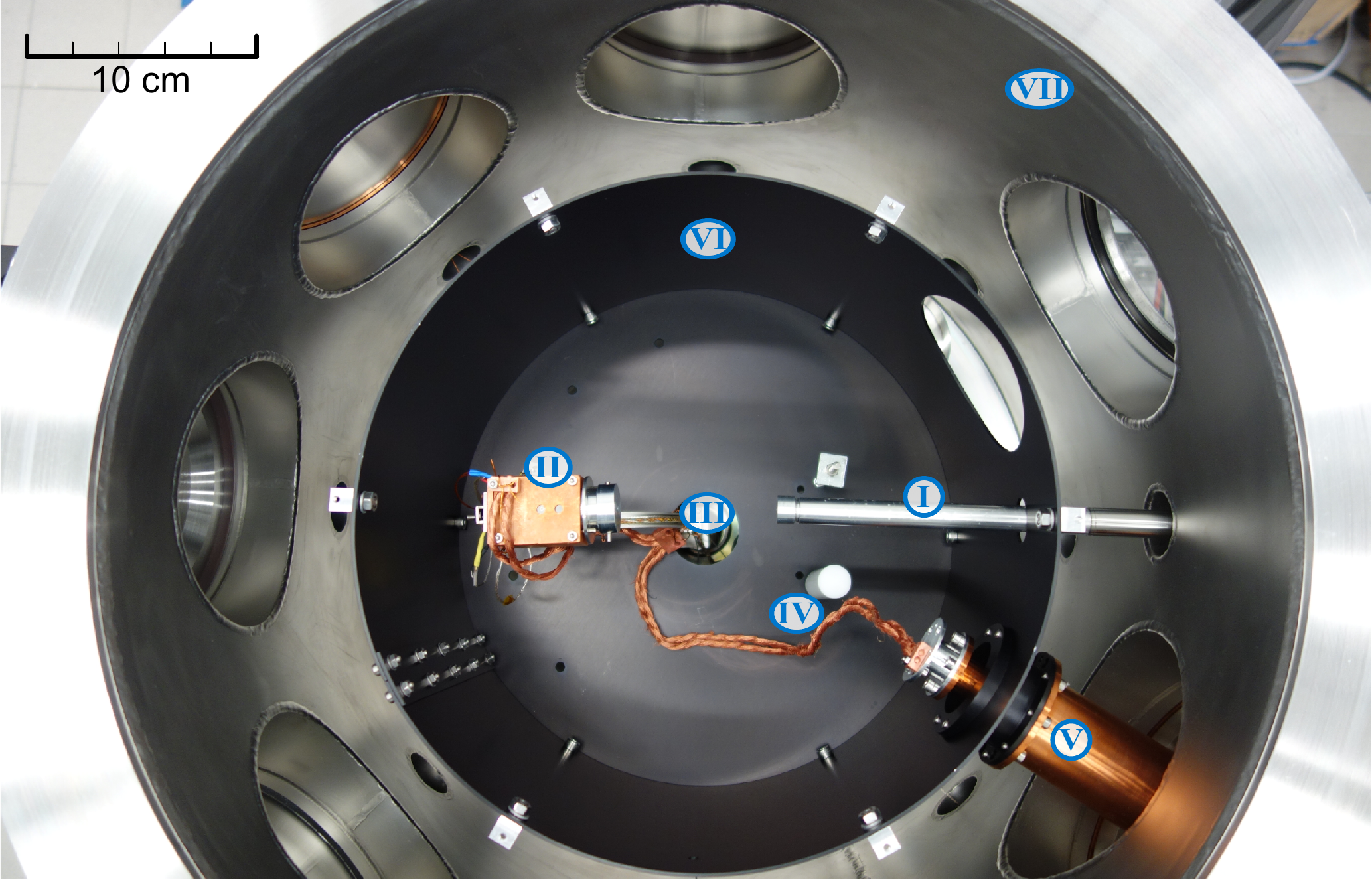}
	\caption[Picture of the main chamber of the setup]{Top view into the main chamber of the reflection setup. The VUV light enters the chamber via a tube and collimator (I) and is monitored by a PMT with a copper housing and an aluminium collimator (II) which is rotated within the plane of this image. The quartz tube filled with the LXe will be located exactly between the collimator and the PMT (III). Stoppers prevent the PMT from colliding with the beam entry (IV). The PMT is cooled via a thermal link to a coldhead (V) and enclosed in a black shield (VI) to reduce thermal irradiation. The entire main chamber is located in a vacuum vessel (VII) and evacuated.}
	\label{fig:Refl-setup-chamber}
\end{figure}

The setup used in this work is located at the University of Münster and was initially designed for the XENON collaboration~\cite{XENON1T17}. It allows us to measure the angular reflectivity for various samples immersed in LXe at wavelengths in the VUV. The xenon used in this work had a natural abundance of isotopes instead of the xenon enriched in \isotope[136]{Xe} that will be used in nEXO~\cite{nEXO18-pCDR}. The individual setup components are described in~\cite{bokeloh13,levy14,wagenpfeil21,althueser20} so only a brief overview is given here. The setup is schematically shown in Figure~\ref{fig:Refl-setup-scheme}. Figures~\ref{fig:Refl-setup-picture} and~\ref{fig:Refl-setup-chamber} show photographs of the setup.

A McPherson 632 deuterium lamp emits a continuous light spectrum above $170\unitnm$. A narrow wavelength band in the VUV is selected by a McPherson 218 plane grating vacuum scanning monochromator. The monochromator output spectrum has been calibrated to provide light around $178\unitnm$ close to the peak LXe scintillation wavelength~\cite{fujii15}. This calibration is described in~\cite{levy14} and is crucial given the strong wavelength dependence of the refractive index of LXe~\cite{hitachi05}. No polarization effect is mentioned by McPherson for the monochromator so the light output is assumed to be randomly polarized. A slit size of $2\unitmm$ is chosen and translates to a width of $5.3\unitnm$ for the monochromator emission spectrum which is a relatively narrow wavelength band compared to the half-height bandwidth of the xenon scintillation spectrum at $10.2\unitnm$~\cite{fujii15}. The VUV photons are then guided through a collection lens and a collimator with a $1.5\unitmm$ diameter aperture before arriving in the main chamber.

The main chamber is shown in Figure~\ref{fig:Refl-setup-chamber} and accommodates the quartz tube with the reflection samples inside and a rotatable photomultiplier tube~(PMT) to register the reflected photons. The custom-made quartz tube is manufactured by VUV-transmissive quartz from \textit{Aachener Quarzglas-Technologie Heinrich} and can be filled with LXe. The tube has a height of $50\unitmm$, an inner diameter of $40\unitmm$ and a wall thickness of $3\unitmm$. It is sealed by a bottom and top stainless steel flange and two Viton O-rings. The flanges are pressed onto the tube by three threaded rods evenly spaced on the outside of the flanges.

The xenon gas enters and exits the quartz tube via two stainless steel ports in the top flange and is liquefied by a cold finger connected to an Iwatani PDC 08 pulse tube refrigerator. The LXe is kept at a constant temperature of around $179\unitK$ and at a pressure of around $2.1\unitbar$.

The collimator, through which the VUV photons enter the main chamber, is firmly attached to the main chamber. The AOI $\theta_{\text{i}}$ is changed by rotating the entire quartz tube including the sample at cold operation. The reflection samples are clamped on a sample holder fixed in the center of the bottom flange, such that the rotation axis of the tube lies in the plane of the sample surface.

VUV photons reflected by the sample in the quartz tube generate an angular distribution of reflected photons in the main chamber. These photons are detected by a UV-sensitive Hamamatsu R8520-406 PMT in counting mode. The PMT is attached to a suspension system connected to a rotational feedthrough at the bottom of the main chamber. A stepper motor rotates the suspension and the PMT around the quartz tube with a step size of $0.45\unitdeg$. This way, a wide angular range can be investigated for each AOI-setting of the sample in the quartz tube. A collimator with a $2\unitmm$ diameter aperture is attached to the front of the PMT and allows a good angular resolution of the angular reflectivity. The PMT is cooled to below $-10\unitC$ to reduce dark noise to a minimum, and wrapped with a mu-metal shield to attenuate the Earth's magnetic field. The PMT count rate is registered via a fast amplifier and a leading edge discriminator.

The main chamber also accommodates a cold shield to reduce thermal radiation. The cold shield is covered with a black coating to reduce photon scattering. A turbomolecular pump evacuates the chamber to $3{\times}10^{-5}\unitmbar$ and thus below the saturation vapour pressure of water at the temperatures of the LXe. The quartz tube and all other components that come in contact with the LXe are cleaned thoroughly and evacuated prior to each xenon fill process.

\subsection{Reflection samples}\label{ssec:samples}

The samples examined in the reflectivity setup are listed in Table~\ref{tab:Refl-samples}. VUV-sensitive SiPMs manufactured by FBK and Hamamatsu are currently considered as photosensor candidates for use in nEXO. The $4^{\text{th}}$ generation VUV-sensitive Multi-Pixel Photon Counter from the S13370 series from Hamamatsu (in the following denoted as VUV4) has been introduced in 2018. FBK provided samples from their VUV-HD LF model fabricated in 2016 and a bare silicon wafer sample covered with a $1.5\unitmum$ thick film of SiO$_2$. The sample examined in~\cite{nEXO19-reflVac} is from the same silicon wafer as the one used in this work. The Large-Area Avalanche Photodiode~(LA-APD) is a mechanically and optically intact spare part from the EXO-200 detector~\cite{EXO11-design}.

\begin{small}
\begin{table}[h!]
	\centering
		\begin{tabular}{lllllll}
		\toprule	
			Sample		& Model						& Manufacturer	& Size											& Pixel Pitch		& FF				& Ref.							\\
		\midrule
			VUV4			& MPPC S13370			& Hamamatsu			& ($6\unitmm$)$^2$					& $50\unitmum$	& $60\,\%$	& \cite{ghassemi17}	\\
			FBK				& VUV-HD 1 (LF)		& FBK						& ($10\unitmm$)$^2$					& $35\unitmum$	& $80\,\%$	& \cite{capasso20}	\\
			Si Wafer	& 								& FBK						& ($20\unitmm$)$^2$					& plain					&						& 									\\
			LA-APD		& SD630-70-75-500	& Adv. Photonix	& \diameter $\,25.6\unitmm$	& plain					& 					& \cite{neilson09}	\\
		\bottomrule
		\end{tabular}
	\caption[Reflectivity samples]{List of the reflection samples examined in this work including the sample size, pixel pitch and the fill factor~(FF) which is the ratio of the active surface area to the total area. References to information by the manufacturers are given if available.}
	\label{tab:Refl-samples}
\end{table}
\end{small}

\subsection{Measurement procedure and data processing}\label{ssec:process}

\begin{figure}[t!]
	\centering
	\includegraphics[width=0.99\textwidth]{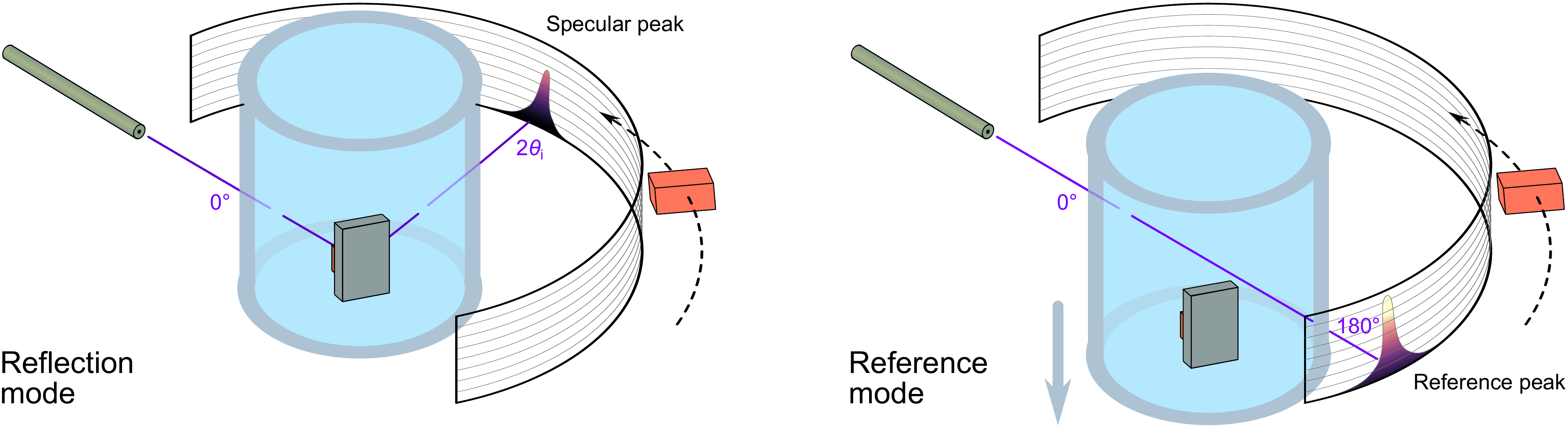}
	\caption[3d scheme of the reflectivity measurement principle]{Schematic overview of the reflectivity measurement principle. The PMT scans the angular distribution of reflected photons in counter-clockwise direction. In reflection mode (left), the VUV beam reflects off the sample in the quartz tube and a specular peak is registered. In reference mode, the quartz tube is lowered and the VUV beam crosses the LXe above the sample holder.}
	\label{fig:Refl-3dscheme}
\end{figure}

The diameter of the VUV beam is approximately $2.9\unitmm$ in vacuum at the position of the PMT. In LXe, the beam is compressed in the plane of reflection due to the large refractive index of LXe. The PMT is rotated around the quartz tube in counter-clockwise direction and monitors the reflected photon rate per angular step. The angular position of the PMT is set to $0\unitdeg$ at the location where the VUV beam enters the quartz tube and $180\unitdeg$ directly opposite of the collimator (see Figure~\ref{fig:Refl-3dscheme}).

In the following, the notation \textit{campaign} is used for a set of measurements with one specific sample. Each campaign consists of two sets of runs: reflection runs and reference runs -- which are both depicted in Figure~\ref{fig:Refl-3dscheme}. Reflection runs are angle-resolved measurements acquired in reflection mode where the VUV beam is at level with the sample. The beam enters the chamber at an angle of $0\unitdeg$, hits the sample under an AOI $\theta_{\text{i}}$ and a specular peak occurs at $2\theta_{\text{i}}$. Reference runs are also angle-resolved measurements -- but acquired by lowering the entire quartz tube until the VUV beam passes the LXe above the sample holder and exits the quartz tube at $180\unitdeg$.

The specular reflectivity $r$ at a specific $\theta_{\text{i}}$ is calculated by dividing the specular peak integral acquired in reflection mode by the reference peak integral acquired in reference mode:
\begin{equation*}
	r\left(\theta_{\text{i}}\right) = \frac{\int\text{Specular peak}}{\int\text{Reference peak}}
\end{equation*}
\noindent
Specular peak integrals are always smaller than reference peak integrals since a fraction of light is absorbed by the sample. The PMT background noise is subtracted from all measurements prior to integration. Reference peaks are acquired several times within a set of reflection runs to monitor the reference peak stability.

The metal lines connecting the SiPM microcells in parallel create a periodic micro-structure on the sensor surface. If these micro-structures have different slopes than that of the microcells, they will create secondary specular peaks, which will be taken into account separately.

The specular peaks are fitted with a dedicated beam shape model to determine the beam position. This model is based on a linear integration of a Gaussian beam along its horizontal plane with the PMT aperture defining the integration limits. The model contains the peak position as fit variable, from which the AOI $\theta_{\text{i}}$ is calculated. The PMT angle scale is calibrated based on the position of the reference peaks which is set to $180\unitdeg$. Detailed descriptions of the measurement proceudre and the beam shape model can be found in~\cite{wagenpfeil21}.

\section{Systematic uncertainties}\label{sec:systematics}

Several systematic uncertainties and their impact on the reflectivity measurement are investigated:

\vspace{\baselineskip}
\noindent
\textbf{Setup alignment}
\newline
\noindent
The PMT used to detect the reflected photons is attached to a four degrees of freedom mechanical feedthrough at the bottom of the main chamber allowing the PMT to rotate as well as to move in all three dimensions. The plane in which the PMT orbits the quartz tube needs to be aligned properly to maximize the light yield and to exclude systematic uncertainties -- e.g.~if the rotation axis of the PMT is shifted relative to the cylinder axis of the quartz tube. In this case, reflected light does not hit the PMT surface perpendicularly, and the monitored photon rate changes with scanning angle. The alignment procedure involves a rough adjustment of the PMT orbit by eye and several runs for fine-tuning. It is repeated after changing the reflection sample. The alignment is confirmed to be successful for all studies presented here. The procedures are explained in depth in~\cite{levy14,wagenpfeil21}. The remaining uncertainty is estimated to be below $1\,\%$. 

\vspace{\baselineskip}
\noindent
\textbf{PMT background rate}
\newline
\noindent
The dark rate of the PMT is measured and subtracted from all measurements. No fluctuations in the dark rate have been observed within timescales of several hours, partly due to the low temperatures within the main chamber. Also, no angular dependency of the dark rate has been observed. All vacuum windows are sealed and all vacuum gauges switched off during data acquisition. A small noise peak was observed at $180\unitdeg$ due to fluorescence relaxations of a $\text{MgF}_2$ window at the monochromator exit. Atmospheric muons triggering scintillation light emission in the LXe were calculated to contribute to the PMT background rate to less than $3\,\%$. A systematic error of $1\,\%$ has been set to account for any remaining background rate artifacts.

\vspace{\baselineskip}
\noindent
\textbf{Light source}
\newline
\noindent
A warm-up time of $1$ hour was set prior to data acquisition to properly warm up the light source. The light source also exhibited slightly different photon fluxes after being switched off and on again. This is taken into account by acquiring independent sets of reference runs for each campaign.

The light source intensity was observed to vary with the laboratory temperature. Alternating runs were acquired in the reference position to closely monitor the reference peak integral. The integrals are interpolated over time and the interpolation is used to determine the reference at the timestamp of each reflection run acquired in-between. The remaining systematic uncertainty for the calculation of the angular reflectivity is estimated to be below $1\,\%$. The same interpolation is performed for the reference peak positions to obtain the reference angle at the reflection run timestamps from which the AOI is determined.

Statistical uncertainties are negligible due to the considerable VUV light flux emitted by the light source which leads to a photon rate of up to $160\unitkHz$ in the reference peak maximum.

\vspace{\baselineskip}
\noindent
\textbf{Bubble formation}
\newline
\noindent
Bubbles in the LXe have been observed to affect the path of the VUV photons via scattering effects.

Streams of small bubbles occurred in the LXe -- especially outside the thermodynamic equilibrium e.g.~after a change of the xenon temperature. The bubbles originate from microscopic defects on the sample or sample holder and block or scatter a fraction of VUV photons crossing the tube above the holder. For all data, the system was operated at temperatures and pressures to avoid bubble formation.

An additional observed phenomenon was one single large bubble floating immediately beneath the LXe surface and blocking the path of the VUV photons. Such bubbles typically grew slowly until they touched the sample holder, burst and reformed. This phenomenon was avoided by extended cleaning procedures of all internal components that are in contact with the xenon.

\vspace{\baselineskip}
\noindent
\textbf{Reference angle stability}
\newline
\noindent
The reference angle facing the entry collimator is determined by the position of the maximum of the reference peaks. It has been observed that the reference angle varies slightly with the rotation of the quartz tube. This variation is caused by a small tilt of the quartz tube of about $6\unitdeg$, leading to a small refraction effect. This translates into an uncertainty of the AOI which is calculated based on the position of the reference peaks. The systematic uncertainty has been determined to be $1.1\unitdeg$ for all measurement campaigns.

\vspace{\baselineskip}
\noindent
\textbf{Rayleigh scattering}
\newline
\noindent
Rayleigh scattering refers to the elastic scattering of photons by particles smaller than the wavelength. The Rayleigh scattering length in LXe is about $40\unitcm$~\cite{grace17, solovov04} but depends strongly on the xenon purity. The xenon used here was cleaned after purchase with a heated getter but not immediately prior to the measurement runs. A decline in xenon purity was avoided by extended cleaning efforts of all parts that come in contact with LXe during operation.

The inner diameter of the quartz tube is $4.0\unitcm$ so about $91\,\%$ of photons pass through the chamber without experiencing Rayleigh scattering. The quartz tube is cylindrical and the photon path is the same for the reference and the reflection mode. Rayleigh scattering affects both modes to the same degree so no systematic impact on the reflectivity is expected. The same holds for possible impurities in the LXe that absorb VUV photons. Nevertheless, a systematic uncertainty due to Rayleigh scattering is considered for the last $2.0\unitcm$ where the VUV photons take different paths depending on the reference/reflection mode. According to above scattering length, $4.9\,\%$ has been set as a systematic error attributed to Rayleigh scattering.

\section{Results}\label{sec:results}

All angle-resolved reflection measurements comprise a prominent specular peak in the PMT response at $2\theta_{\text{i}}$. No diffuse reflection component has been observed for any sample. Secondary peaks are observed for samples with surface structure.

\subsection{Wafer sample and beam profile}\label{ssec:profile}

\begin{figure}[t!]
	\centering
	\includegraphics[width=0.80\textwidth]{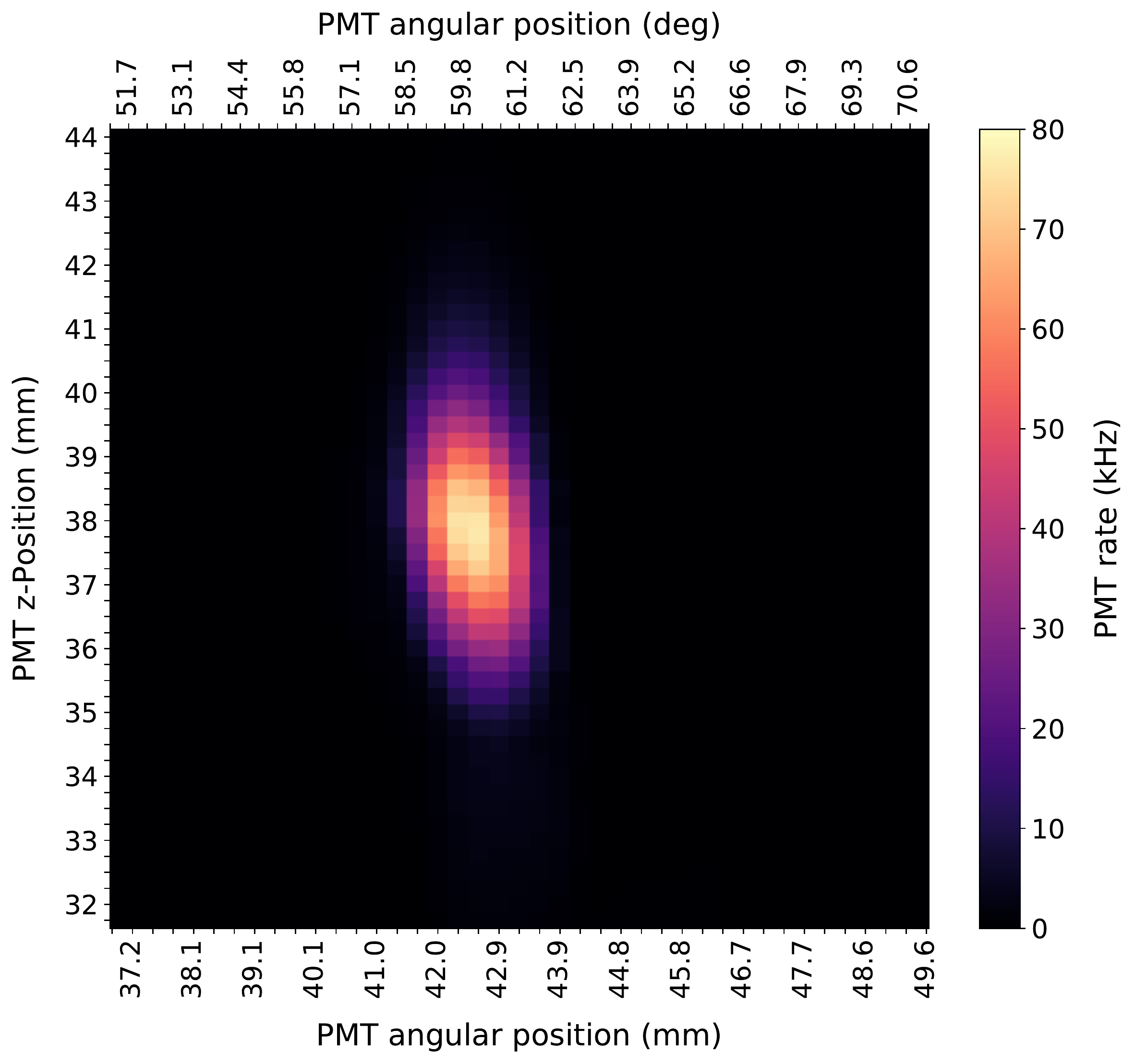}
	\caption[VUV beam profile with Wafer sample under 30 deg AOI]{VUV beam profile acquired with the wafer sample at an AOI of $30\unitdeg$ in LXe. The angular position of the PMT is converted to circle segments on the lower x-axis. The y-axis shows the z-position of the PMT orbit. The beam is compressed in the plane of incidence due to focusing effect caused by the change in refractive index between vacuum, quartz and LXe.}
	\label{fig:Refl-profile}
\end{figure}

The bare silicon wafer sample provided by FBK is coated with $1.5\unitmum$ SiO$_2$. The wafer is the same material from which FBK manufactures their VUV-sensitive SiPMs, but does not have any surface micro-structures. The sample shows a completely smooth surface and is suitable for use as a VUV mirror. Samples from the same wafer material have already been examined in vacuum~\cite{nEXO19-reflVac}.

The VUV beam profile has been studied using the wafer sample in LXe at an AOI of $30\unitdeg$. The profile is acquired by scanning the reflected beam at different z-positions of the PMT orbit and stacking the individual angular measurements. The 2d-projection is shown in Figure~\ref{fig:Refl-profile}.

Due to the large refractive index of the LXe and the shape of the quartz tube, a strong focusing effect acts on the VUV beam and leads to a compression in the plane of incidence. The focal point is located roughly halfway between the tube and the PMT aperture. The beam is uncompressed in the z-direction. The fit model introduced in Section~\ref{ssec:process} is extended to 2d beam profiles and used to extract the horizontal and axial beam width (see~\cite{wagenpfeil21} for details). The vertical width acquired from the fit agrees with previous measurements at the same setup~\cite{bokeloh13}. The compression is modelled with a scaling factor that only acts on the horizontal width and only depends on the known geometry of the quartz tube as well as on the refractive index of LXe, $n_{\text{LXe}}$. The fit yields a value of $1.567{\pm}0.008$ for $n_{\text{LXe}}$, where only the fit uncertainty is quoted here. This value lies within the range of experimental measurements of the refractive index of LXe reaching from $1.5655$ to $1.69$~\cite{chepel94,barkov96,grace17,solovov04}.

The beam profile is slightly inclined due to the tilt of the quartz tube. The tilt angle has been determined to be $(6.02{\pm}0.03)\unitdeg$. The effect of this tilt is taken into account by the systematic error on the final AOI as discussed in Section~\ref{sec:systematics}.

\subsection{Wafer sample reflectivity}\label{ssec:Wafer}

The wafer reflectivity was measured in LXe by acquiring several angular measurements in reflection position as well as in reference position to monitor the beam stability. The reflection angles range from $17\unitdeg$ to $87\unitdeg$. Measurements at smaller AOIs are not possible in the setup since the VUV beam collimator blocks the PMT housing.

\begin{figure}[t!]
	\centering
	\includegraphics[width=0.78\textwidth]{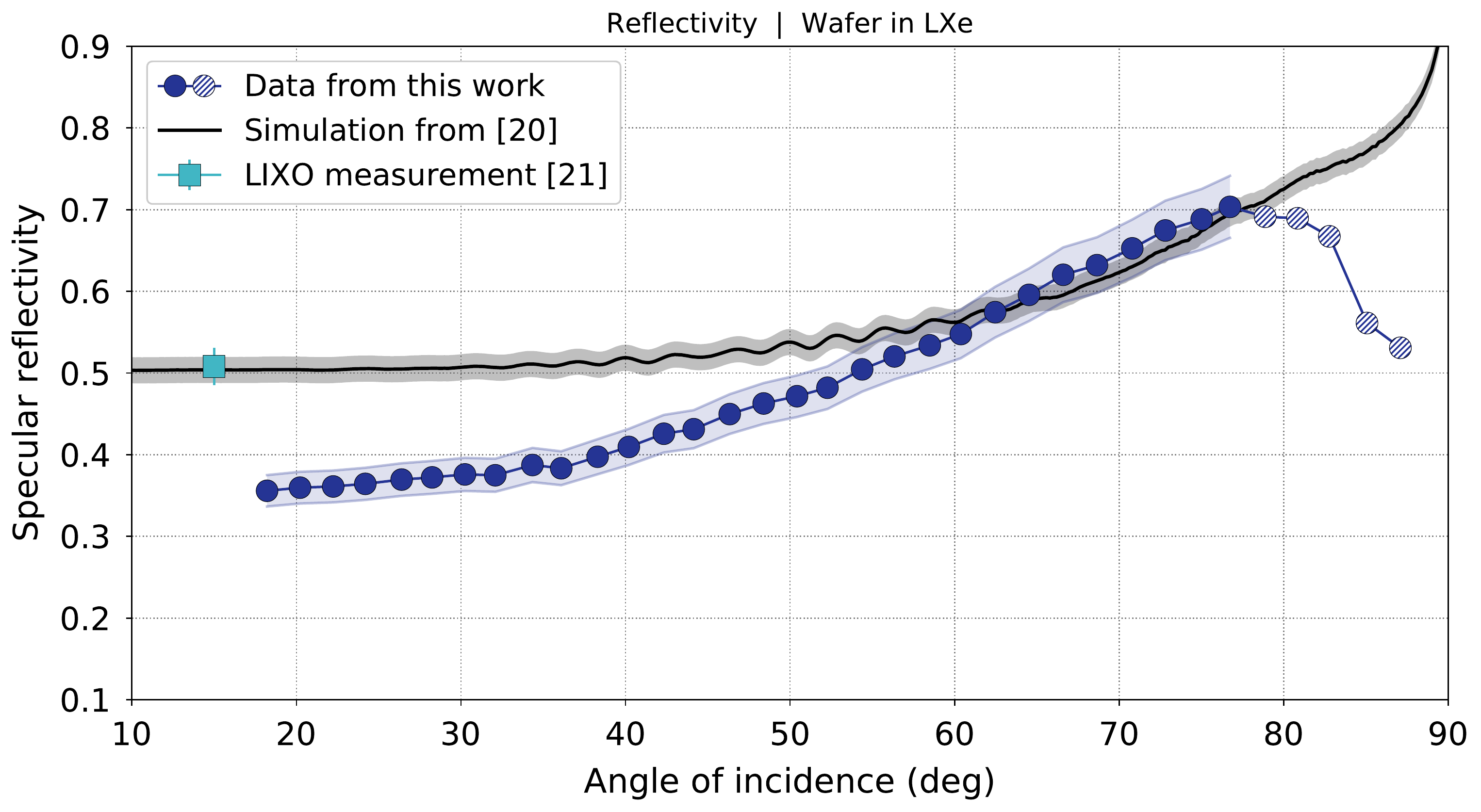}
	\caption[Reflectivity of the wafer sample in vacuum and LXe]{Reflectivity over AOI for the wafer sample in LXe. The uncertainty band includes the contributions discussed in Section~\ref{sec:systematics}. The simulation is adapted from~\cite{nEXO19-reflVac} with the refractive index changed to $n_{\text{LXe}} = 1.57$ according to the above result. Shaded points above $77\unitdeg$ are affected by beam containment issues.}
	\label{fig:Refl-wafer-refl}
\end{figure}

The reference peak integrals are stable to $<1\,\%$ throughout the acquisition time of both campaigns. The reference peaks in LXe are significantly narrower than those of the vacuum campaign due to the LXe focusing. Data is only acquired for a limited angle range around the specular peak (in reflection mode) or around $180\unitdeg$ (in reference mode). Additional measurements scanning the full angle range in reflection mode showed no signature of a diffuse reflection component.

The specular reflectivity is calculated as the ratio of the specular peak integrals from the reflection runs and the reference peak integrals obtained in reference mode. The results are plotted in Figure~\ref{fig:Refl-wafer-refl}. The error band represents the cumulative systematic uncertainty. The individual contributions are discussed above. The reflectivity increases with AOI up to about $77\unitdeg$. Shaded points above this angle indicate that the projected surface of the sample becomes significantly smaller than the width of the VUV beam which is no longer contained on the sample surface. A beam containment correction is applied and demonstrates to be efficient except for the far end of the AOI range examined in this campaign. The drop is a geometric artifact that prevents any valid reflectivity measurements above roughly $77\unitdeg$.

The reflectivity reported here differ from the measurement from LIXO~\cite{nEXO19-reflLXe} -- namely a reflectivity of $50.8{\pm}2.3\,\%$ at an AOI of $15\unitdeg$ determined for a sample from the same wafer. The LIXO measurement is consistent with the prediction for LXe from~\cite{nEXO19-reflVac}. Their simulation is plotted in Figure~\ref{fig:Refl-wafer-refl} with the refractive index $n_{\text{LXe}}$ changed from $1.69$ to $1.57$. The simulation does not fully agree with the data in this work. This can be due to difficulties with modelling samples in LXe based on optical parameters obtained in studies in vacuum. Alternatively, this may point to systematic uncertainties of this work -- especially at AOIs below $53\unitdeg$ -- that could not be resolved. Further investigations are warranted to resolve the discrepancy. In particular, a wafer measurement in the same setup in vacuum may provide additional information without the ambiguity of the $n_{\text{LXe}}$ value.

\subsection{FBK VUV-HD reflectivity}\label{ssec:FBK}

\begin{figure}[t!]
	\centering
	\includegraphics[width=0.84\textwidth]{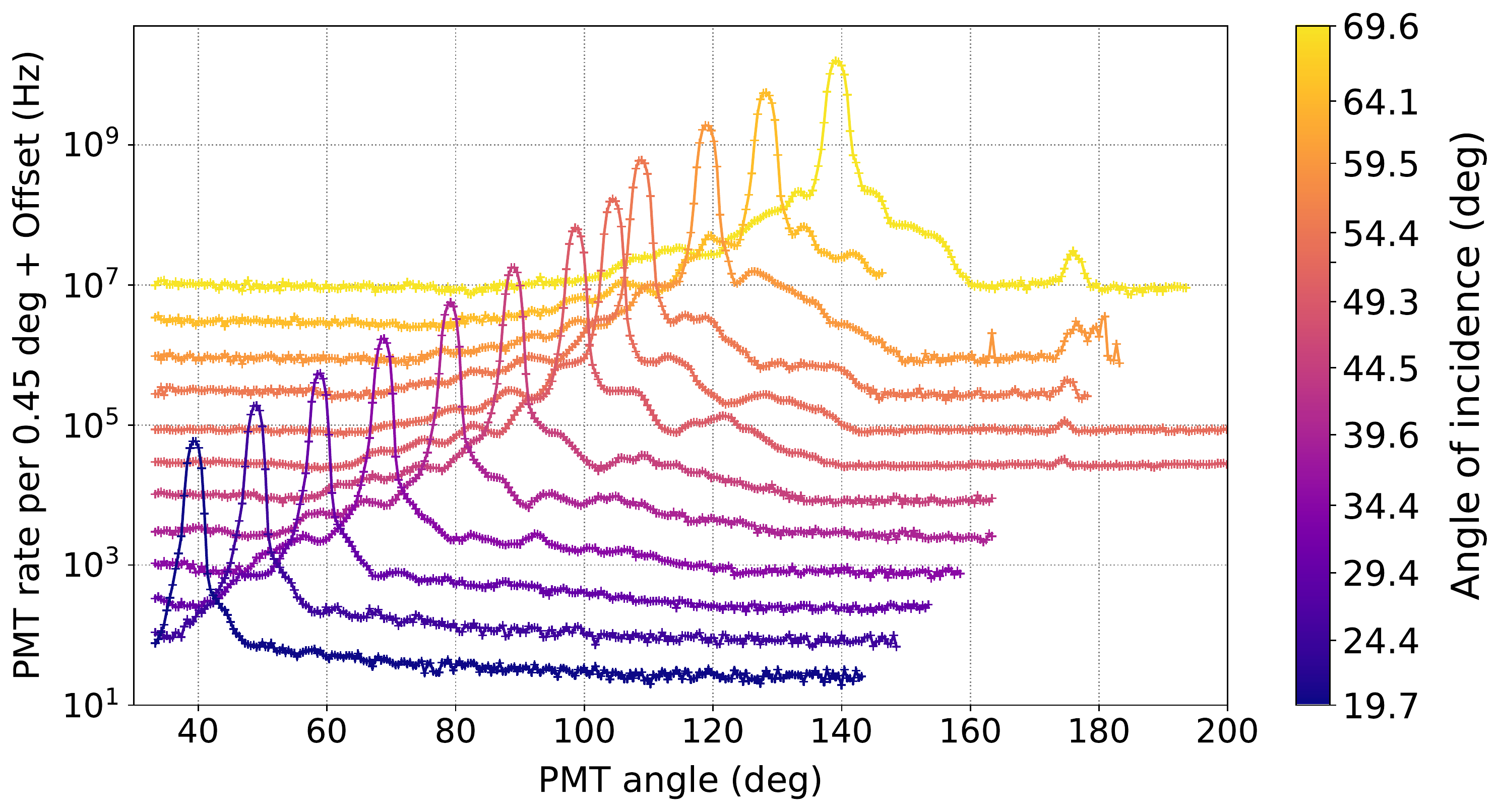}
	\caption[Reflection measurements of the full angle FBK campaign]{Angular distribution of reflected photons measured in the full angle FBK VUV-HD campaign in LXe. The distinct specular peaks are accompanied by small secondary structures due to the micro-structure on top of the SiPM surface. The curves are plotted with an equidistant offset of $\sqrt{10}$ for better visualization.}
	\label{fig:Refl-FBK-rflx}
\end{figure}

The VUV-HD SiPM from FBK is manufactured from the same wafer material examined above and includes micro-structures on its surface.

Two reflectivity campaigns have been performed in LXe. For the first campaign, only a small angle range around the specular peak was scanned to provide a large number of AOI settings. For the second campaign, the full angular range of the PMT is used which allows to investigate possible additional features in the photon distribution other than the specular peak.

Figure~\ref{fig:Refl-FBK-rflx} shows the set of reflection runs from the second, full-angle campaign. The color indicates the AOI setting of the quartz tube. Each run contains one prominent specular peak, while additional secondary structures are observed due to the surface micro-structure. 

The specular reflectivity is again calculated by dividing the integrals of the specular peaks by the ones of the reference peaks. Both campaigns agree within their systematic uncertainties. At larger AOIs -- depending on the size of the sample -- the effective sample surface in the beam path becomes smaller than the beam diameter and a beam containment correction is applied. This correction is based on the integration of a 2d Gaussian beam with rectangular integration limits defined by the sample size, where the horizontal length decreases with AOI via a cosine~\cite{wagenpfeil21}.

\begin{figure}[t!]
	\centering
	\includegraphics[width=0.80\textwidth]{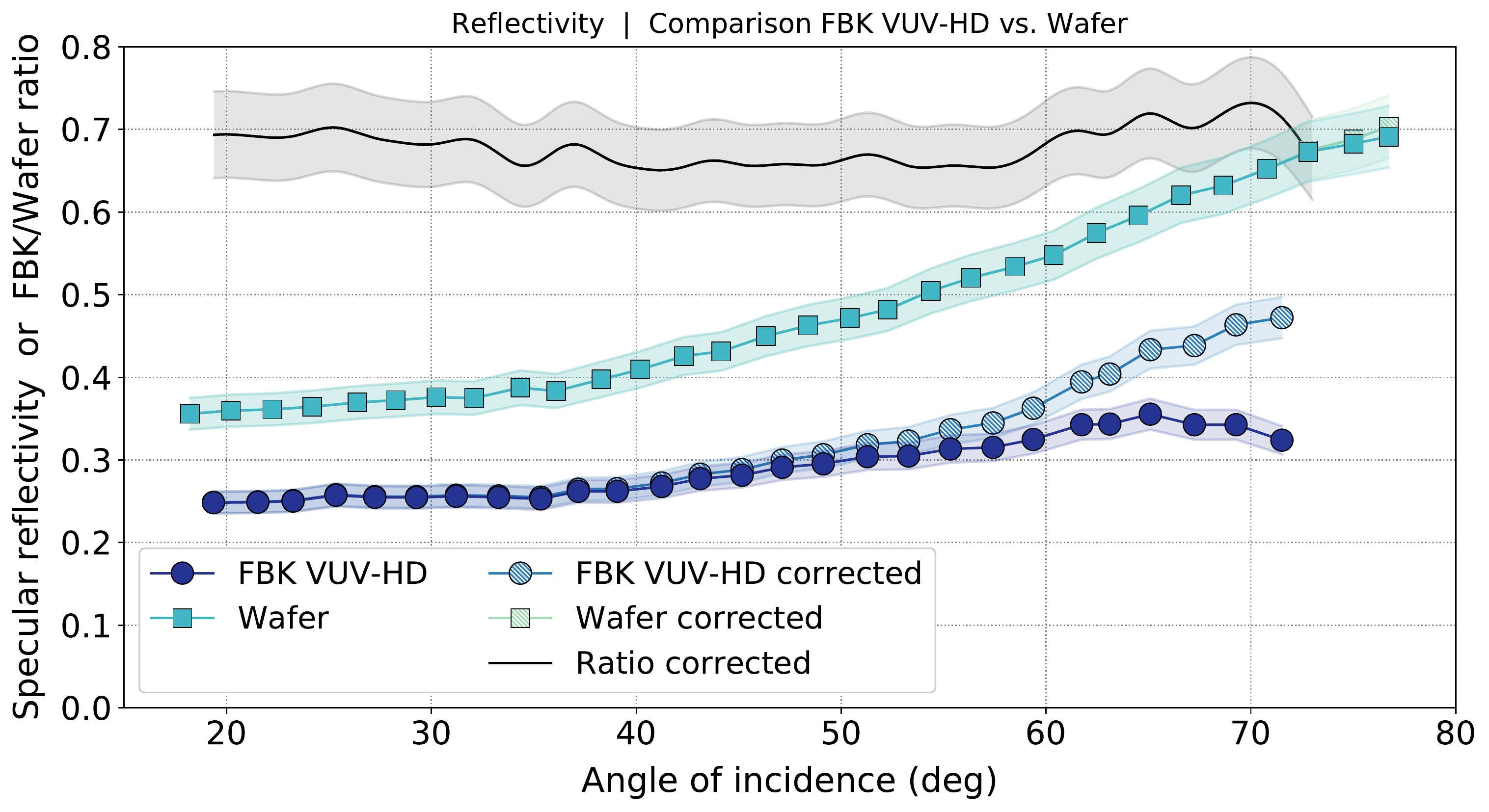}
	\caption[Reflectivity of the FBK sample compared to the wafer]{Reflectivity vs.~AOI of the FBK VUV-HD sample compared to the LXe wafer campaign. Both data sets are shown with and without beam containment correction. The black line shows the ratio between both corrected reflectivity datasets.}
	\label{fig:Refl-FBK-refl}
\end{figure}

Figure~\ref{fig:Refl-FBK-refl} shows the reflectivity of both the FBK VUV-HD and the wafer sample in LXe. The general shape of both reflectivity curves is similar since the FBK VUV-HD is manufactured from the same wafer samples. The FBK VUV-HD has an overall smaller reflectivity compared to the wafer sample. This was expected since a certain fraction of incident photons are absorbed or reflected elsewhere by the surface micro-structure which the wafer sample lacks. The ratio of both corrected reflectivity curves is therefore constant with respect to the AOI as shown in Figure~\ref{fig:Refl-FBK-refl} with a value of about $69\,\%$. The value differs from the fill factor reported in Table~\ref{tab:Refl-samples} since it depends on the surface geometry while the fill factor describes the ratio of photoactive to full surface.

\begin{figure}[t]
	\vspace{5mm}
	\centering
	\begin{minipage}[t]{.48\textwidth}
 		 \centering
 		 \includegraphics[width=0.99\textwidth]{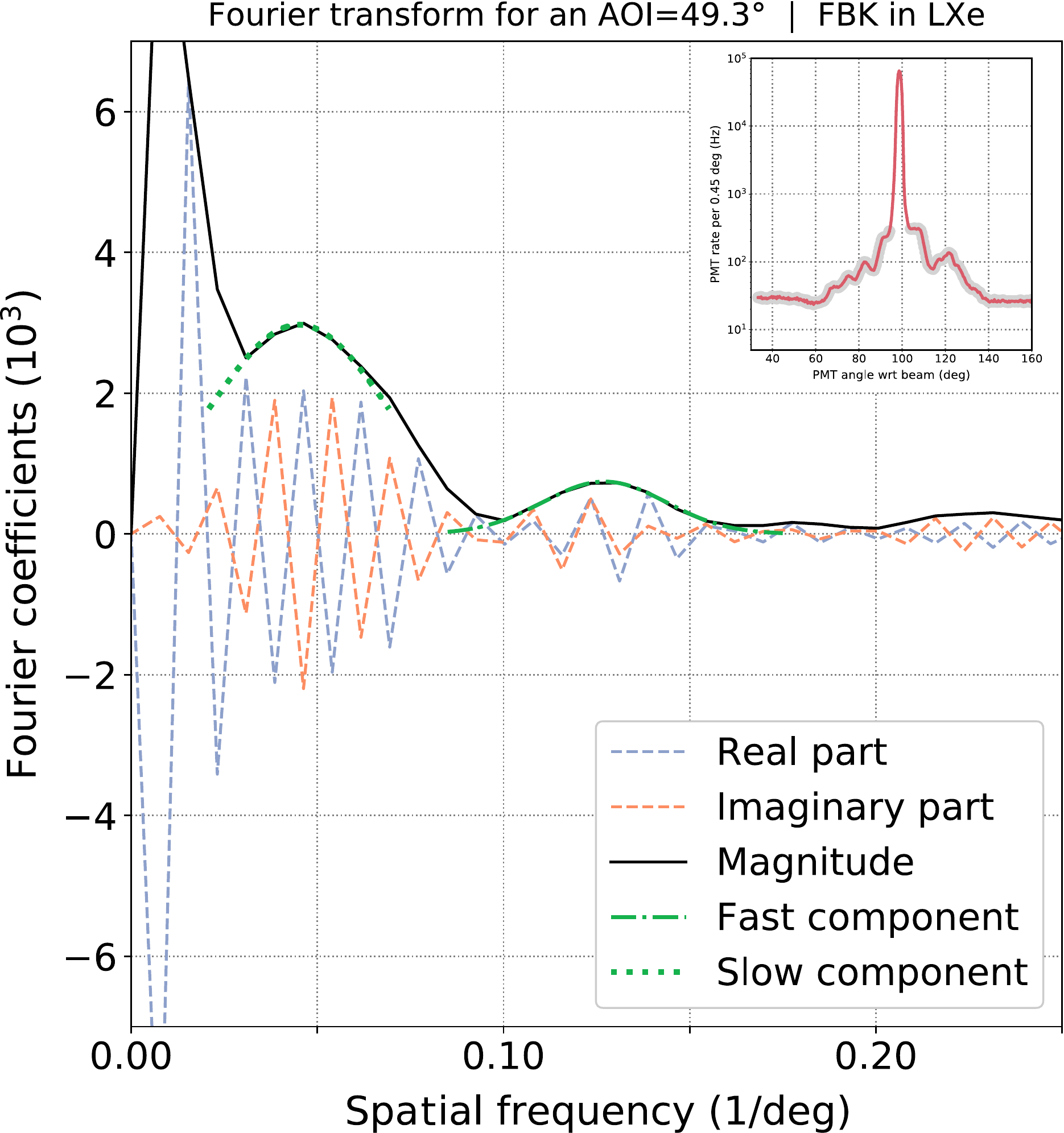}
	\end{minipage}
	\hfill
	\begin{minipage}[t]{.48\textwidth}
 		 \centering
 		\includegraphics[width=0.99\textwidth]{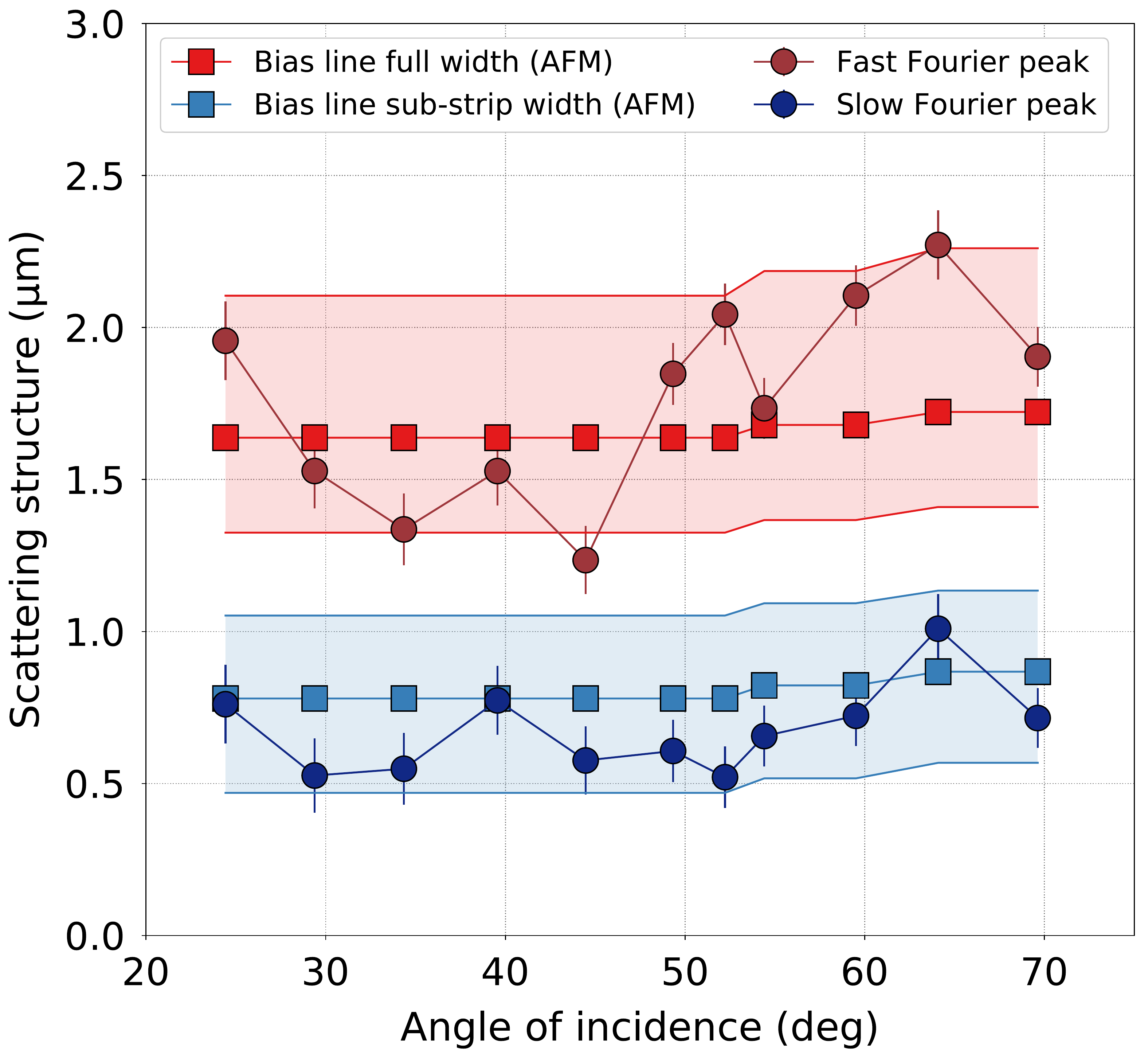}
	\end{minipage}
	\caption[FBK Fourier spectrum at $49.3\unitdeg$ and FBK scattering structures]{\textbf{Left:} Fourier coefficients of a spatial transformation of the FBK VUV-HD reflectivity measurement at $49.3\unitdeg$ from Figure~\ref{fig:Refl-FBK-rflx}. Only the greyish underlaid data of the angular reflectivity is processed (see inlet). Two peaks emerge: a slow component around $0.05\unitpdeg$ and a fast one around $0.12\unitpdeg$ and analysed with a Gaussian fit. \textbf{Right:} Lateral size of the scattering structure on the FBK VUV-HD surface reconstructed from the slow (blue) and a fast (red) components (dots). Squares show the geometric size of the bias lines (red) and bias sub-strips (blue) determined with AFM height profiles.}
	\label{fig:Refl-FBK-structure}
\end{figure}

Oscillations can be observed in the FBK VUV-HD reflection runs in Figure~\ref{fig:Refl-FBK-rflx} due to interference caused by the periodicity of the micro-structure on the SiPM surface. The oscillation is investigated via a Fourier analysis. An example is shown in the left in Figure~\ref{fig:Refl-FBK-structure} for a reflection run at an AOI of $49.3\unitdeg$. The Fourier transform yields a spatial frequency spectrum of the surface structure responsible for the interference pattern and allow to calculate the lateral dimension of this structure -- in this case the bias lines. Two dominant periodicities emerge in the Fourier spectra for all runs in Figure~\ref{fig:Refl-FBK-rflx}. The fast component was found to agree very well with the dimension of the micro-structure bias lines of roughly $1.6\unitmum$. The slow component fits to the width of the two sub-strips that form the bias lines (see Figure~\ref{fig:Refl-FBK-structure}, right). The micro-structure dimensions were determined via atomic force microscopy~(AFM) height profiles of the SiPM surface and are plotted as squares in Figure~\ref{fig:Refl-FBK-structure}. The corresponding uncertainty bands account for the inclination of the edges of the bias lines, i.e.~the bottom width of the bias lines gives an upper limit of the periodic structure while the top width gives a lower limit.

The magnitude of the oscillations is very small and has no effect on the angular reflectivity and the optical properties of nEXO. The findings concerning the influence of the micro-structure are discussed in~\cite{wagenpfeil21} in detail.

\subsection{Hamamatsu VUV4}\label{ssec:VUV4}

\begin{figure}[t!]
	\centering
	\includegraphics[width=0.92\textwidth]{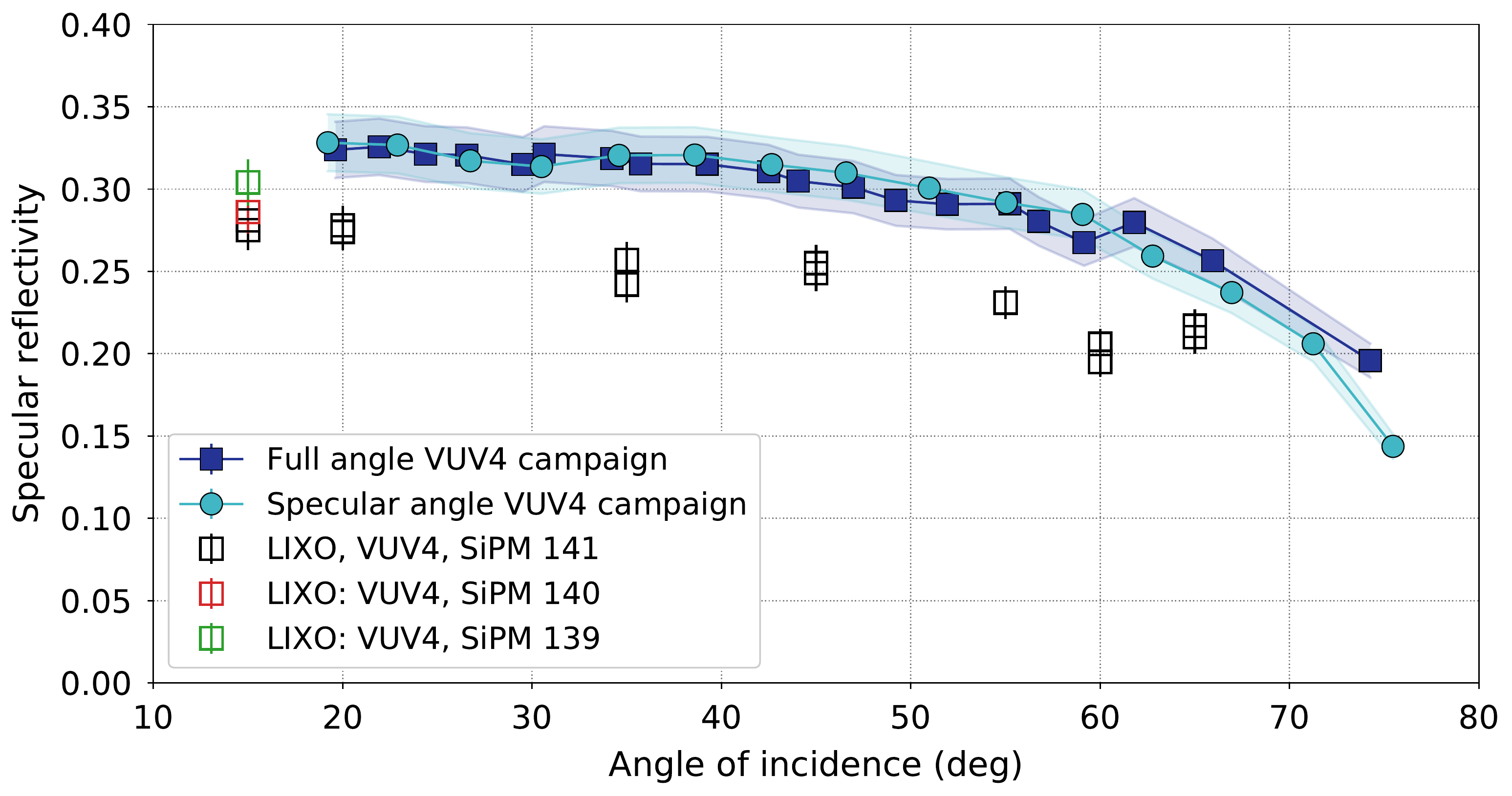}
	\caption[Reflectivity of the VUV4 sample in LXe compared with literature]{Reflectivity over AOI of the VUV4 sample in LXe. The reflectivity is shown for both VUV4 campaigns and compared to results with three VUV4 samples reported by the LIXO group in~\cite{nEXO19-reflLXe}.}
	\label{fig:Refl-VUV4-refl}
\end{figure}

The VUV4 has been examined with one full angle and several specular angle campaigns in LXe. In the specular angle campaigns, a range of only ${\pm}10\unitdeg$ around the position of the specular peak is scanned and integrated. In the full angle campaign, the PMT is rotated all the way around the sample but the same angle range as in the specular campaigns is used to integrate the specular peaks and calculate the specular reflectivity plotted in Figure~\ref{fig:Refl-VUV4-refl}.

The specular reflectivity derived from all campaigns is consistent within the systematic uncertainties over the entire AOI range. Unlike the other samples examined in this work, the VUV4 reflectivity is obsevered to decrease with AOI. This is consistent with the specular reflectivity measurements in~\cite{nEXO19-reflLXe}, where this behaviour was observed first. The reflectivity of SiPM 141 derived in~\cite{nEXO19-reflLXe} is about $19\,\%$~(relative) smaller than what is reported here. If the uncertainties of~\cite{nEXO19-reflLXe} and this work are combined with the measurement range accomplished at $\theta_\text{i}=15\unitdeg$ with all three SiPMs from~\cite{nEXO19-reflLXe}, the inconsistency of both studies nearly vanishes. The difference in specular reflectivity may be caused by device-by-device fluctuations, handling practices or humidity exposure.

As in the case of the FBK VUV-HD, the micro-structure of the VUV4 causes additional features in the reflection runs. The surface structure is significantly more complex for the VUV4. Additionally, the bias lines are inclined to the microcell surface with a predominant slope of ${\pm}10\unitdeg$ according to AFM scans. This causes two distinct secondary peaks in the reflection runs roughly $20\unitdeg$ left and right of the main specular peak. The additional peaks can be interpreted as additonal specular peak caused by the inclined bias lines which contributed to the total reflectivity of the SiPM. The integration of these additional peaks yields the additional number of photons reflected under different angles compared to the specular angle from the microcell surface. As a result, the total reflectivity of the VUV4 is up to $12.8\,\%$ (relative) larger compared to the pure specular reflectivity. Details of these results can be found in~\cite{wagenpfeil21}.

\subsection{EXO-200 LA-APD}\label{ssec:APD}

An optically intact spare part of the large-area avalanche photodiodes (LA-APD) used in the EXO-200 detector~\cite{EXO11-design} has been investigated in LXe. The data is included in the summary plot in Figure~\ref{fig:Refl-all}.

\section{Discussion}\label{sec:conclusion}

\begin{figure}[t!]
	\centering
	\includegraphics[width=0.80\textwidth]{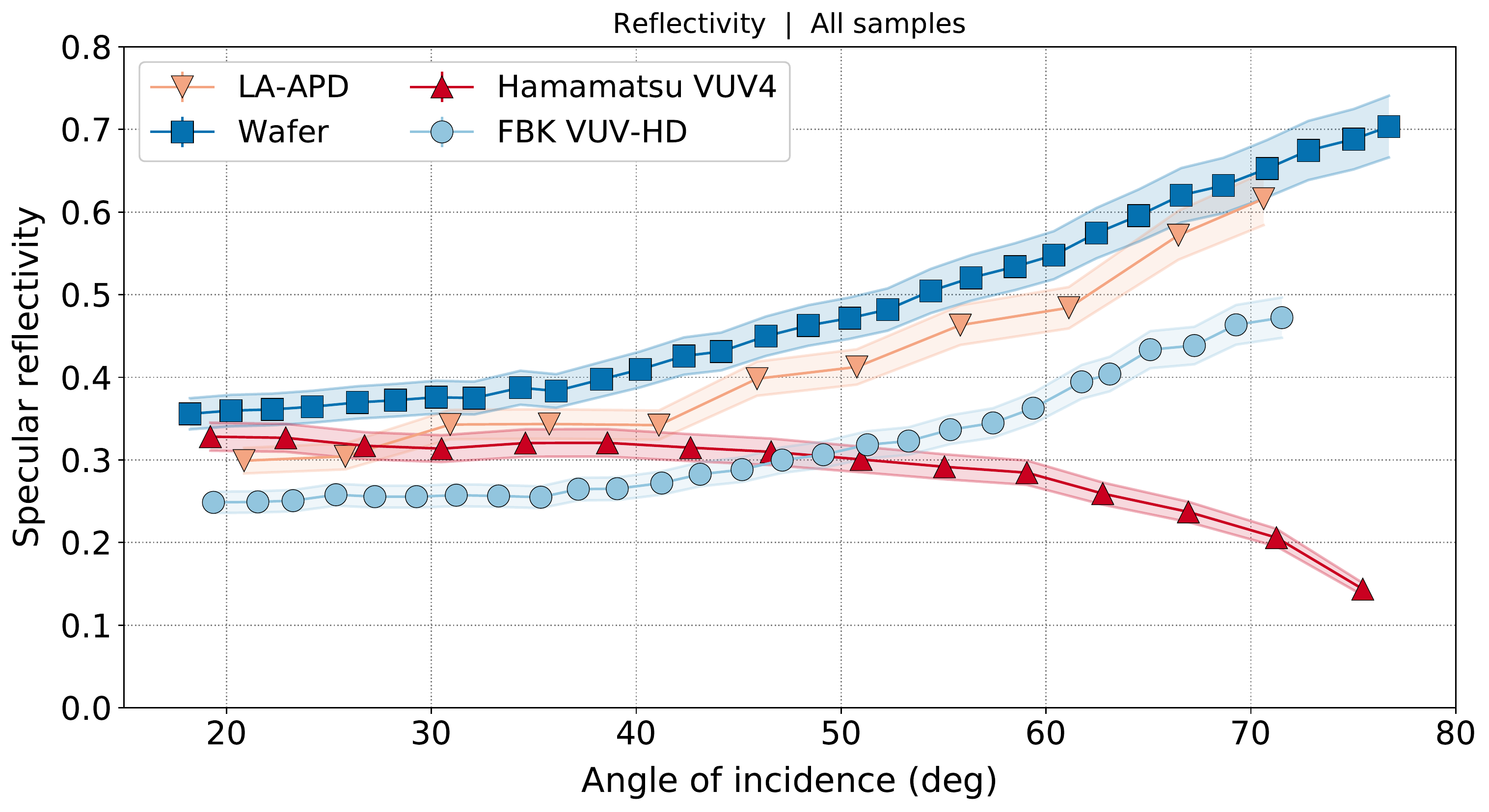}
	\caption[Overview of all reflectivity studies]{Reflectivity in LXe over AOI for all four samples examined in this work. \textit{FBK VUV-HD} and \textit{Hamamatsu VUV4} refer to the corresponding VUV-sensitive SiPM samples and \textit{Wafer} to the wafer sample from FBK. A spare \textit{LA-APD} from the EXO-200 experiment was also investigated. The uncertainty bands are based on the discussion in Section~\ref{sec:systematics}.}
	\label{fig:Refl-all}
\end{figure}

Figure~\ref{fig:Refl-all} shows the reflectivity in LXe of all samples examined in this work. The reflectivity has been measured to increase with AOI for all samples except the VUV4 SiPMs. The results presented in this work are currently used in the light response simulation of the nEXO detector and the impact of the SiPM reflectivity will be discussed in the upcoming sensitivity publication by the nEXO collaboration.

This study presents the first measurements of the FBK VUV-HD and LA-APD reflectivities in LXe. The VUV4 reflectivity follows the same decreasing behavior with AOI first observed in~\cite{nEXO19-reflLXe} while remaining differences are consistent with device-by-device fluctuations and uncertainties concerning the sample surface. This observation underlines the importance of documenting the storage and exposure history of all samples as a possible cause of systematic offsets.

As for the wafer sample, the disagreement between the measurements and the simulation in~\cite{nEXO19-reflVac} suggest that the optical properties of a given SiPM -- i.e.~the refractive index $n$ and the extinction coefficient $k$ -- cannot be measured in vacuum and applied for LXe scintillation wavelengths using simple extrapolations. Such simulations are highly sensitive to the exact photon spectrum used in reflectivity setups due to the strong wavelength-dependence of $n_{\text{LXe}}$ (see Figure~8 in~\cite{nEXO19-reflVac}). On the other hand, the experimental reflectivity of the wafer sample determined in~\cite{nEXO19-reflLXe} is in contradiction to the results presented here. Further systematic studies of the wafer reflectivity in vacuum and LXe are therefore advisable.


%


\acknowledgments

Support for nEXO comes from the Office of Nuclear Physics of the Department of Energy and NSF in the United States, from NSERC, CFI, FRQNT, NRC, and the McDonald Institute (CFREF) in Canada, from IBS in Korea, from RFBR (18-02-00550) in Russia, and from CAS and NSFC in China.


\end{document}